\documentclass
[aps,pre,showpacs,twocolumn,superscriptaddress,unsortedaddress]{revtex4}%
\usepackage{bm}
\usepackage{graphicx}
\usepackage{amsmath}
\usepackage{amsfonts}
\usepackage{amssymb}%
\bibstyle{approve.bib}

\begin{document}
\title{Calculation of the elastic scattering properties in a\ \ ultra-cold Fermi-Bose
and Bose-Bose Rb-K vapor}
\author{M. Kemal \"{O}zt\"{u}rk}
\email{ozturkm@gazi.edu.tr}
\author{S\"{u}leyman \"{O}z\c{c}elik}
\email{suleyman@gazi.edu.tr}
\affiliation{Physics Department, University of Gazi, 06500 Teknikokullar, Ankara, Turkiye}
\date{Today}

\begin{abstract}
The calculation of the elastic scattering properties of mixtures composed by
rubidium and potassium atoms is reported and compared with experimental
results in detail. The improved potentials for both molecular states the
singlet x$^{1}%
{\textstyle\sum}
$ $_{g}^{+}$ and the triplet a$^{3}%
{\textstyle\sum}
$ $_{u}^{+}$ of the RbK are presented, and the scattering lengths $a_{t}$ and
the effective range $r_{e}$ are calculated with the help of this potential
using WKB and Numerov methods for Rb-K in the triplet and singlet state. In
addition, the convergence of these scattering properties as the depending on a
$K_{0}$ parameter is investigated using Quantum Defect Theory. The evaporative
cooling other results that include the cross section and the spin-charge cross
section, the rate coefficient as a function of the energy and the ultra-low
temperature are also presented in this study.

\end{abstract}
\maketitle

\section{\textbf{INTRODUCTION}}

Collisions of atoms at ultracold temperatures have received considerable
attention because of their importance in the cooling and trapping of atoms
\cite{1}and molecules \cite{2} and Bose Einstein condensation by Sympathetic
cooling \cite{3}. Ultracold mixtures of alkali metal atoms have recently been
the subject of intensive experimental research, therefore the interest in the
detailed knowledge of ultracold mixtures of alkali atoms has theoretically
increased \cite{4,5}. Even, while collisions either between alkali often same
species or between different isotopes have often been well characterized
\cite{6}, collision between different alkali species have a still intensive
exploring field. The condensed atoms and molecules may form a two-species
Bose-Einstein condensate in which a new class of coherent phenomena is
predicted to occur \cite{7}.

In an ultracold mixture, the interaction between the components is described
in terms of a very important parameter, the interspecies s-wave scattering
length. The character of the interaction between different alkali species in
low-temperatures is determined by signs ($\pm)$ and magnitude of the
scattering length $a$: for bosonic atom with $a$%
$>$%
0 Bose condensate is stable, for $a$%
$<$%
0 it is unstable \cite{8}. This parameter also determines the efficiency of
the sympathetic cooling procedure \cite{3}. In addition, knowledge of the
scattering length is important in order to predict macroscopic properties of
these mixed systems, such as the stability and phase separation of binary
degenerate systems, or microscopic properties, such as the occurrence of
resonances and the cross section and the spin-charge cross section in the
interspecies collisions.

Accurate calculations of s-wave scattering length and effective range for
diatomic potentials are important due to the elastic collisions in these
temperatures dominated s-wave scattering. In recent years, these quantities
have been calculated by several theoretical methods \cite{9,10,11,12,13}. The
accurate determination of scattering length depends on choosing the
interaction potentials. Collision processes at near- zero temperatures
($\sim\mu$K) are sensitive to the details of the interaction potentials
between the colliding systems.

Previous theoretical and experimental calculations of the scattering length
for Rb+K interaction have appeared in the literatures \cite{3,14}. The purpose
of this paper, we are to calculate scattering lengths, effective ranges and
phase shifts-cross section for the collisions of Rb+K at ultra-cold
temperatures with the Semiclassical method (WKB) \cite{15} and also solve the
Schr\"{o}dinger equation using Numerov method numerically \cite{16}. We have
investigated the convergence of some scattering properties obtained with help
of these methods as the dependence of a $K^{0}$ parameter using Quantum Defect
Theory \cite{13}. The interaction of rubidium-potassium atoms in states
($^{1}\sum_{g}^{+})$ and ($^{3}\sum_{u}^{+})$ is described by short and long
range potentials which includes the cutoff functions introduced to truncate
the unphysical short-range contribution of the polarization potential near the
origin\cite{17}. Cutoff functions depending on the cutoff radius are adjusted
by comparing the expressed potential form to the curve calculated with
\textit{ab-initio} by Rousseau et.al. \cite{18}. Some results we have found
are compared with theoretical values in the ref. \cite{14} and the
experimental result \cite{3}.

This paper is designed as the follow form, we discuss the computational
process in Sec. 2. The interaction potentials, which play a crucial role in
evaporate cooling and also determine many properties of the condensate, are
given in Sec. 3. The ultracold scattering properties(the scattering length,
effective range, phase shifts and the rate coefficients) are computed for
collisions of the Rb-K isotop systems in the singlet and triplet states at
ultra-cold temperatures in Sec. 4 and 5, respectively, and the convergences of
the scattering range and the effective range are also investigated in detail.
We conclude with brief conclusions.

\section{\textbf{PROCESSES}}

The scattering length is defined from the asymptotic behavior of the solution
of the radial Schr\"{o}dinger equation at zero energy:%

\begin{equation}
\left[  {\frac{d^{2}}{dr^{2}}-2\mu V(r)}^{{2S_{AB}+1}}{+k^{2}+\frac
{l(l+1)}{r^{2}}}\right]  y_{l}(r)=0,\quad\quad\quad\label{eq1}%
\end{equation}
in atomic unit. Here ${\mu}$ is the reduced mass of the atomic system and
${V(r)}^{{2S_{AB}+1}}$is the molecular potential between two atoms. Asymptotic
behavior of the wave function is%

\begin{equation}
y(r)=s{\ }\,r+s_{0} \label{eq2}%
\end{equation}
as $r\rightarrow\infty$, where $s$ and $s_{0}$ are constants. The scattering
length is given by%

\begin{equation}
a=-\frac{s}{s_{0}}, \label{eq3}%
\end{equation}
these constants are obtained by G. F. Gribakin using et al. using WKB method
from semiclassical behavior and with help of the exact solution at zero-energy
\cite{19}. Eq. (\ref{eq3}) is transformed to the form%

\begin{equation}
a=\bar{a}\left\{  {1-\tan\{\varphi-\pi/8\}}\right\}  \label{eq4}%
\end{equation}
where $\bar{a}$ is the \textquotedblleft mean\textquotedblright\ or
\textquotedblleft typical\textquotedblright\ scattering length determined by
the asymptotic behavior of the potential through the parameter $\gamma
=\sqrt{2\mu\kappa}$ ($\mu$ is reduced mass, and $\kappa=c_{6}$ is Van der
Waals constant) for atom-atom interactions:%

\begin{equation}
\bar{a}=\frac{\Gamma(3/4)}{\Gamma(1/4)}\sqrt{2\gamma} \label{eq5}%
\end{equation}

Scattering length also depends a semiclassical phase $\varphi$ calculated at
zero energy from classical turning point $r_{0} $ where $y(r_{0} ) = 0$, to infinity,%

\begin{equation}
\varphi=\int\limits_{r_{0}}^{\infty}{\sqrt{-2\mu{V(r)}^{{2S_{AB}+1}}}dr}.
\label{eq6}%
\end{equation}

The total number of vibrational levels with zero orbital angular momentum was
derived from $\varphi$, $N_{bd}=\{\varphi/\pi-(n-1)/2(n-2)\}+1$, where
$\mathit{n}$ is are given by integer value 6 for interaction of mixing or same
species atoms \cite{19}. A similar calculation of the semiclassical phase at
negative energies, $E=-1/(2\mu a^{2})$ yields $\varphi(E)=\varphi-\sqrt{\pi
}\gamma^{2/n}a^{-(n-2)/n}\Gamma(\frac{1}{2}+1/n)/[(n-2)\Gamma(1+1/n)].$It is
often used for studying quantization of vibrational levels near the
dissociation limit in diatomic molecules\cite{20}.

The effective range $r_{e}$ can be written in terms of zero-energy solutions
of the partial-wave equation if $y_{0}(r)$ is the solution of the partial-wave
equation at $k=0$ for zero-potential, and normalized as%

\begin{equation}
y_{0}=\frac{\sin(kr+\delta_{0})}{\sin\delta_{0}} \label{eq7}%
\end{equation}
as $k\rightarrow0$, and if $y_{0}(r)$ is the normalized solution of the radial
equation (\ref{eq1}) for the s-partial wave at zero energy (for non-zero
potential) as $y(r)\sim y_{0}(r)$ at $r\rightarrow\infty$. Then, effective
range can be written as \cite{15,21}%

\begin{equation}
r_{e}=2\int\limits_{0}^{\infty}\{y{{_{0}^{2}(r)-y}}^{{2}}{{(r)\}}dr}.
\label{eq8}%
\end{equation}

This integral converges provided $y(r)$ approaches $y_{0}(r)$ rapidly enough
as $r\rightarrow\infty$. This requires $V(r)$ to decrease faster than $r^{-5}%
$. For $-c_{6}/r^{n}$potential type, this expression is adjusted by Flambaum
et al.\cite{22} using WKB method and the exact solution of the radial equation
at zero--energy as%

\begin{equation}
r_{e}=\frac{\sqrt{2\gamma}}{3}\left[  {\frac{\Gamma\left(  {1/4}\right)
}{\Gamma(3/4)}-2\frac{\sqrt{2\gamma}}{\mathit{a}}+\frac{\Gamma(3/4)}%
{\Gamma(1/4)}\frac{4\gamma}{a^{2}}}\right]  . \label{eq9}%
\end{equation}

The mean effective range may be easily obtained by replacing eq.(\ref{eq4})
into eq. (\ref{eq9}) as%

\begin{equation}
\overline{r_{e}}=\frac{\sqrt{\gamma}}{3}\left[
\begin{array}
[c]{c}%
{\frac{\sqrt{2}\Gamma\left(  {1/4}\right)  }{\Gamma(3/4)}-\frac{4\sqrt{\gamma
}}{\overline{\mathit{a}}(1+\tan(\frac{1}{8}(\pi-8\varphi)))}}\\
\text{ \ \ \ \ \ \ \ \ \ }{+\frac{\Gamma(3/4)}{\Gamma(1/4)(1+\tan(\frac{1}%
{8}(\pi-8\varphi)))^{2}}\frac{4\sqrt{2}\gamma}{\overline{\mathit{a}}^{2}}}%
\end{array}
\right]  . \label{eq10}%
\end{equation}

The low-energy scattering is dominated by the contribution $l=0$. At values of
$k$ close to zero, the $l=0$ phase shifts $\delta_{0}$ can be represented by a
power series expansion in $k$ \cite{23}%

\begin{equation}
\label{eq11}k\cot\delta_{0} = - \frac{1}{a} + \frac{1}{2}r_{e} k^{2} +
o\left(  {k^{3}} \right)  .
\end{equation}

This expression is known as Bethe formula.

\section{\textbf{INTERACTION POTENTIALS}}

The interaction between the atoms has two features: First, the potential at
large distances behave as an inverse power of the interatomic distance,
$V(r)=-\kappa/r^{n}$ with $n=6$ for spherically symmetric two atoms in their
ground state. The asymptotic parameter $\kappa\equiv c_{6}$ is known quite
well for most atomic pairs of interest. Second, for atoms other than hydrogen
and helium the potential curve is usually quite deep and also the
electron-exchange part of the atomic interaction is repulsive at smaller
distance as for the singlet and triplet terms of alkali- metal atoms.
\textquotedblleft Deep\textquotedblright\ here means that the wave functions
of the atomic pair oscillate many times within the potential well, even at
very low collision energies. The interatomic potential supports a large number
of vibration levels. This latter feature enables one to use the semiclassical
approximation to describe the motion of atoms within potential well \cite{22}.

\begin{figure}
[ptb]
\begin{center}
\includegraphics[
height=3.8761in,
width=3.6434in
]%
{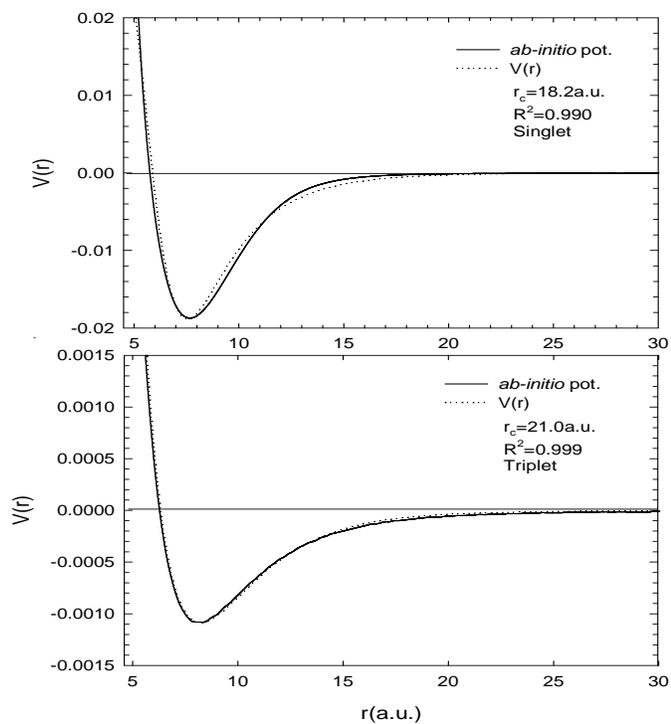}%
\caption{Potential energy (dotted line) curves obtained as compared with
experimental (solid line) potential curve for the best cutoff radii in the
singlet and triplet states of the moleculer KRb. The insets show $R^{2}$
values and cutoff radii}%
\label{fig1}%
\end{center}
\end{figure}

The mixing alkali-metal atoms have unpaired electron spin $s=1/2$ in the
ground state Therefore, the interaction between two different alkali atoms in
the ground state results in the formation of a diatomic molecule, which is
described by two terms corresponding to the total spin $s_{ab}=s_{a}+s_{b}$ of
the system equals to 1 or 0. The state with $s_{ab}=0$ and $1$ corresponds to
the singlet state and triplet state, respectively. The probability of the spin
exchange is determined by both the dependence of potential curves on the
internuclear distance $r$ and the splitting between singlet and triplet terms.
The exchange interaction is analytically calculated using the surface integral
method by Smirnov and Chibisov \cite{24}, which yields%

\begin{equation}
v_{exc.}(r)=\frac{1}{2}(v_{u}-v_{g})=r^{\alpha}e^{(-\beta{\ }r)}J.
\label{eq12}%
\end{equation}
Where $\alpha,\beta$ and $C$ are constants. In this equation, $J$ \ is given
by $\sum\nolimits_{n}\mathit{f}{_{n}J_{n}r^{n}}$, here, $f_{n}$ are fit
constants adjusted by fit program for the best R square in this paper, and
$J_{n}$ are the integral constants. For Rb-K, these values of constants taken
from the paper of Smirnov and Chibisov are given in Table \ref{tab1}. The
long-range part of the interaction potential is given by%

\begin{equation}
v_{long-range}=-%
{\displaystyle\sum\nolimits_{m}^{even}}
\frac{C_{m}}{r^{m}}f_{c}(r), \label{eq13}%
\end{equation}
this term is the sum of Van der Waals terms, multiplied by the relevant cutoff
function $f_{c}(r)$ to cancel $1/r^{n}$ divergence at small distances. The
area under r axis of the interaction potential is very important to specify
the ultracold scattering properties, and its value is calculated from the eq.
(\ref{eq6}). For any different two potential, when it has same semiclassical
phase shifts, the scattering properties aren't affected from the change of
other parameters of potential such as the depth of potential \cite{25}. It is
the fact that theoretical potential should actually adopt with \textit{ab-
initio} one. Therefore, in the singlet state, the some sensitive Wan der Waals
parameters taken from ref. \cite{26} was multiplied with fit parameter
$C_{f}^{c}\,$. As shown in the Fig.%
\ref{fig1}%
, it concludes that the potentials are excellent adapt due to R square, and we
may also name the $C_{f}^{c}\,$as the corrected constant for Wan der Waals
parameters in singlet and triplet state, and this technique may be useful for
other studies. $C_{6}\,$, $C_{8}\,$and $C_{10}$ parameters were taken from A.
Derevianko et al.'s paper (Ref. \cite{26}). The cutoff function is analogous
to that used for the H-H${\ }^{3}\Sigma_{u}^{+}$ potential \cite{27}.%

\begin{equation}
\label{eq14}f_{c} (r) = \xi\left(  {r - r_{c} } \right)  + \xi(r_{c} - r)e^{ -
(\frac{r_{c} }{r} - 1)^{2}},
\end{equation}

Where $\xi(z)$ is unit step function $\xi(z)=1(0)$, when $r>(<)0$. There, the
only free parameter for the potential sum is the cutoff radius $r_{c},$
governing the decrease of the function $f_{c}(r)$ for $r<r_{c}$. The potential
of the collisions Rb+K consists of above the expressed two terms , the sum of
Eqs. (\ref{eq12}) and (\ref{eq13}), is given as%

\begin{equation}
\label{eq15}V^{2s_{AB} + 1}(r) = V(r)_{long - range} + ( - 1)^{s_{ab}
}V(r)_{exc.}%
\end{equation}

The magnitude of the cutoff parameters $r_{c}$ has been adjusted by comparing
this potential with the experimental curve in triplet state obtained by
Rousseau et.al. \cite{18} using \textit{ab-initio} for the large range.

The most relevant potential curves for different r$_{c}$ values are
represented in Fig.\ref{fig1}, together with the experimental calculation and
their R square values (inset). Owing to R square value, there is in excellent
agreement between experimental and theoretical curves for the cutoff radius
determined from the fit curve.

\section{\textbf{THE CONVERGENGE AND CALCULATION SCATTERING LENGTH AND
EFFECTIVE RANGE OF }$\mathbf{Rb+K}$}

In this section, the scattering length and the effective range for the
collisions of Rb+K in singlet and triplet state at ultra-cold temperature have
been calculated with semiclassical method (WKB) \cite{15} and also determined
by solving the Schr\"{o}dinger equation by Numerov method \cite{16} for the
potential given with Eq.(\ref{eq15}). Also, we investigated the convergence of
these scattering properties as a function of energy parameter($K^{0})$ using
Quantum Defect Theory\cite{13}.

In third column of Table \ref{tab2}, the results of the scattering length
calculated from Eq.(\ref{eq4}) with help of the WKB method are represented for
the collision of mixtures composed by rubidium and potassium atoms in singlet
and triplet state described with the potentials in eq.(\ref{eq15})
characterized by the term of cutoff radius. Some parameters are important in
the determination of $a(a.u.)$ for WKB method. We also obtain these
parameters, the mean scattering length $\bar{a}$ and the zero-energy
semiclassical phases $\varphi$, given in Table \ref{tab3} using values of
asymptotic parameter $\gamma$. Here, the mean scattering length with
asymptotic behavior of the potential in eq.(\ref{eq15}) has been calculated
from equation (\ref{eq5}).

For calculation numerically scattering length , we obtained s-wave phase shift
$\delta_{0}$ using the potentials in eq.(\ref{eq15}), then we do it by solving
radial equation using Numerov algorithm at $e=k^{2}/2\mu$ in atomic unit and
finding $\delta_{0}$ from the asymptotic behavior of the wave function
$y(r)=\sin(kr+\delta_{0})$. The phases at small momenta $k$ are used to
extract the scattering length numerically from $a=-\lim_{k\longrightarrow
0}\left(  d\delta_{0}/dk\right)  $ The scattering lengths obtained via this
method are given in the fourth columns of Table \ref{tab2}. The results
obtained in two methods are compared with values found in the literature .

From the measured thermalization rate, using the model presented in
Ref.\cite{28}, the large value for the zero-energy s-wave scattering length
are obtained using conventional double magneto-optical trap (MOT) apparatus
\cite{29,30}. G. Modugno, G Ferrari and et. al have estimated a direct
measurement of the scattering lengths for the $^{41}$K-$^{87}$Rb pair
\cite{3}. They also deduced the collisional behavior of other isotope
combinations by the help of mass scaling in the experimental columns of Table
\ref{tab2}. \cite{14}. Then, for the $^{40}$K-$^{87}$Rb and$^{41}$K-$^{87}$Rb
pairs, Simoni et al. investigated the scattering properties in the presence of
several magnetic field \cite{31}.

As shown in the Table \ref{tab2}, the scattering lengths calculated for the
collisions of rubidium- potassium pairs is in excellently agreement with the
experimental scattering length when $V(r)$ potential has the excellent adopt
with \textit{ab initio} curve due to the best R square value.

Even our results shows that the stability of large condensates requires
repulsive interactions (positive $a)$, whereas for attractive interactions
(negative $a)$ it is unstable, only a finite number of atoms can be found in
condensate state in a trap. As shown Tablo \ref{tab2}, the scattering lengths
found as negative value due to changes in the $V(r),$ lead to a condensate
triplet state where the number of atoms is limited to a small critical value
determined by the magnitude of $a$. In contrast, we have observed the positive
scattering lengths that produce stable condensates. Also, from the scattering
length given as a function of the semiclassical phase shift in the eq.
(\ref{eq4}), we say that the minor changes in phase shifts have strongly
affected the scattering amplitude. It concludes that the ultra-cold scattering
depends on the potential changes given by phase. It must be remarked that our
emphasis for $^{41}$K-$^{87}$Rb pair is weakly dependent on the number
$N_{bd}$ of bound states supported by \textit{ab-initio} potential. The
numbers of bound states $N_{bd}$ have been computed in the intervals 108-110
and 32-33 for both singlet and triplet states, respectively. The numbers of
bound states of the singlet state is consistent with the result ($N_{b}=32$)
obtained ref.\cite{14} In addition, the mass scalling from one isotope to the
others depends more strongly on $n_{s}$. The calculations in this paper shows
that due to being boson-fermion, of great interest also is the $^{40}$%
K-$^{87}$Rb pair, for which we find a negative and probably large interspecies
scattering length agreed with other result in Table \ref{tab2} which implies
an efficient sympathetic cooling of the fermionic species down to the
degenerate regime \cite{14}.

Effective ranges calculated using Numerov and WKB methods in the function of
phase for the triplet rubidium-Potassium potentials are presented in Table
\ref{tab4} together the calculation of semiclassical phase shifts. The close
agreement between the calculations of $r_{e}$ from Eqs.(\ref{eq8}) and
(\ref{eq11}) confirms the accuracy of the numerical integration of the
partial-wave equation. The size of the scattering lengths and effective ranges
is closely related to the position of the last vibration bound states of the
energy curves, as can be anticipation by inspection of Eq.(\ref{eq3}) and
number of vibrational levels $n_{s}(\varphi/\pi)$ which, consistent with
Levinson's theorem, show that as the binding energy of the highest level tends
to zero, the scattering length tends to $\pm$infinity \cite{15}.

Analytical calculations of scattering lengths are important in investigation
of convergence of the scattering length and the effective range. In many
approaches used to solve collision problems in atomic physics, the tree
dimensional configuration space is divided into two regions separated by a
spherical shell (a core boundary) of radius $\rho$. In the inner region (r%
$<$
$\rho)$ the short-range interaction between two colliding particles is very
complicated and a scattering equation must be solved independently for each
combination of particles. In contrast, it $\rho$ is chosen sufficiently large
the scattering problems in the outer region(r%
$>$
$\rho)$ may be reduced to potential scattering with the long-range potential
accurately approximated by simple analytical expression. A numerical solution
in this region is usually easily approachable \cite{32}. One of these methods
is Quantum Defect Theory described by Gao \cite{13}.

Quantum defect theory of atomic collisions is presented by Bo Gao. Based on
the exact solutions of the Schr\"{o}dinger equation for an attractive
$1/r^{6}$ potential, the theory provides a systematic interpretation of
molecular bound states and atom-atom scattering properties and establishes the
relationship between them. Bo gao finds a definition for the scattering length
and the effective range and s-wave at zero energy as%

\begin{equation}
\left.  {%
\begin{array}
[c]{l}%
a_{l=0}=\frac{2\pi}{[\Gamma(1/4)]^{2}}\frac{K^{0}-1}{K^{0}}\beta_{6}\\
\\
r_{l=0}=\frac{[\Gamma(1/4)]^{2}}{3\pi}\frac{\left[  {K^{0}}\right]  ^{2}%
+1}{\left[  {K^{0}}\right]  ^{2}}\beta_{6}\\
\end{array}
}\right\}  \label{eq16}%
\end{equation}

where $\beta_{6}=\left(  {2\mu\,C_{6}}\right)  ^{1/4}$ and $K^{0}$is the
analytic function of energy.

\begin{figure}
[ptbh]
\begin{center}
\includegraphics[
height=3.8614in,
width=6.0667in
]%
{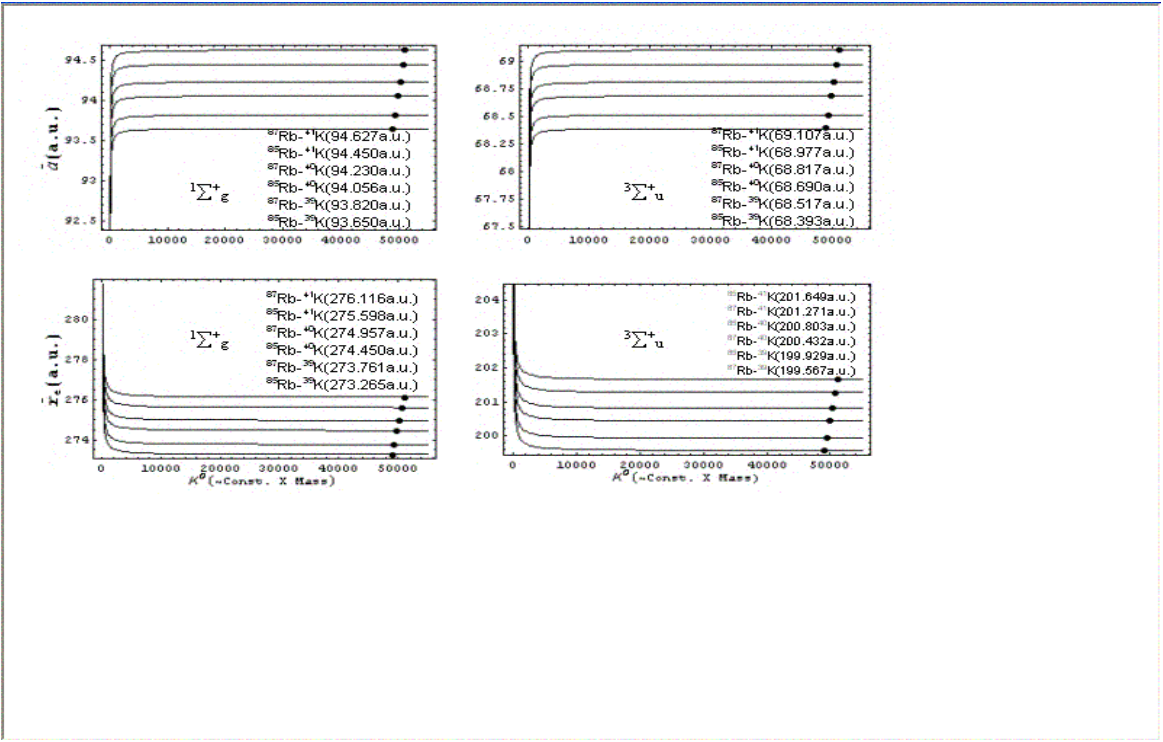}%
\caption{Convergence of the scattering length and the effective range
calculated with QDT as a function of $K$ parameter when $K$ goes infinity.}%
\label{fig2}%
\end{center}
\end{figure}

Figure \ref{fig2} show the scattering length and the effective range curve
plotted with the help of eqs. (\ref{eq16}) and values(dotted points) obtained
by the mean WKB formulas, eqs.(\ref{eq5}) and (\ref{eq10}).

Our graphical representations show that as $K^{0}$ goes infinity, the QDT
results converges excellently to values obtained by WKB method in Fig.
\ref{fig2} . Even, it may be said that the different calculating of the
scattering properties in these two method gives similar results for any
ultracold atom-atom collision (e.g. ref. \cite{33}).

\section{\textbf{PHASE SHIFTS AND CROSS SECTION}}

\begin{figure}
[ptbh]
\begin{center}
\includegraphics[
height=2.9187in,
width=4.0465in
]%
{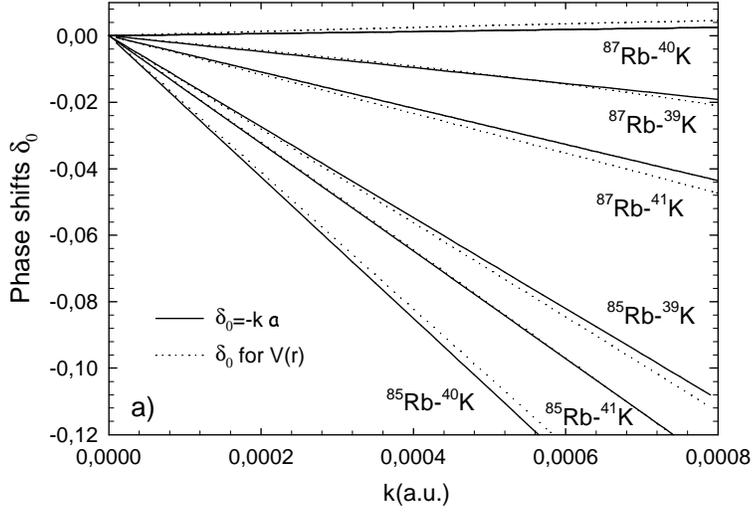}%
\caption{s-wave phase shifts calculated for the $V(r)$ potential in the
singlet and triplet states (Solid lines). Dotted lines show the curves
calculated with $\delta_{0}$= -$a$ $k$.}%
\label{fig3}%
\end{center}
\end{figure}

Scattering length describes behavior of the atom scattering at low energy, and
phase shifts are important parameter for it. Furthermore, at a low energy, E,
of relative motion the phase-shift is $N_{bd}\pi+\delta$, where $N_{bd}$ is
the number of bound states. The phase shift can be calculated by fitting the
solution $y(r)$ to $\sin(kr)$ and $\cos(kr)$, the asymptotic solutions of Eq.
(\ref{eq1}). Let $S_{i}$ and $C_{i}$ denote $\sin(kr_{i})$ and $\cos(kr_{i})$,
respectively. Fitting at $r_{i}$ and $r_{j}$ we find, for a small shift
\cite{15,16},%

\begin{equation}
\delta\approx\tan(\delta)=-\frac{S_{i}y_{j}-S_{j}y_{i}}{C_{i}y_{j}-C_{j}y_{i}%
}\quad, \label{eq17}%
\end{equation}
where $r_{i,j}=r_{0}+(i\,,j)h$ (h is the step-length ). we show the numerical
phase shifts (module $\pi$) in Fig. \ref{fig3} it is also known in more tex
book that the low energy scattering is always dominated by the $l=0$ partial
wave the corresponding phase shift being given by%

\begin{equation}
\delta_{0}=-ka \label{eq18}%
\end{equation}
which is shown for all Rb+K collisions in Fig. \ref{fig3}. Here, to plot phase
shifts we used the semiclassical scattering lengths. As momenta goes zero, for
$l=0$, the s-phase shifts are very close in the singlet state. it concludes
that when a gas mixture composed of fermion-boson or boson-boson is cooled
below a critical temperature, the mixing atoms condensates in the lowest
quantum state. Moreover, if the atoms are bosons, a cloud of atoms occupies
the same quantum state at low temperature. If the atoms are fermions, cooling
gradually brings the gas closer to being a `Fermi sea' in which exactly one
atom occupies each low-energy state \cite{34}.

As shown in fig. \ref{fig3}, their scattering lengths are unstable against
collapse except the $^{40}$K-$^{85}$Rb. For these mixtures of composed by
fermion-boson, the condensates are more attractive than that of boson-boson as
reported in Table \ref{tab3}. In this case, the fermionic cloud cannot shrink
below a certain size determined by the Pauli Exclusion Principle due to
bosonic cloudy \cite{34}.

Elastic singlet and triplet cross sections between two atoms, can be defined by%

\begin{equation}
\sigma_{el.}^{S,T}=\frac{4\pi}{k^{2}}\sum\limits_{l=0}^{\infty}{(2l+1)\sin
^{2}\delta_{l}^{S,T}}, \label{eq19}%
\end{equation}
where $S$ and $T$ stand for singlet and triplet, respectively. We have
obtained the phase shifts ($\delta_{l}$) and cross sections ($d\sigma_{l}$)
for the collision $^{85}$Rb+$^{40}$K with $l=2,4$ values of the angular
momentum and different energies using Numerov method. They are listed in Table
\ref{tab4}. this table shows that the $\delta_{l}$ and $d\sigma_{l}$ for very
low energy have large values lead to the being of more collisions. In the
ultracold gas of the mixing alkali metal atoms dominated by the $l=0$
contribution, the cross sections are given by $\sigma_{el.}^{S,T}%
=8\pi\,a_{S,T}^{2}$ and $\sigma_{el.}^{S,T}=4\pi\,a_{S,T}^{2}$ for boson-boson
systems and boson-fermion systems, respectively \cite{35}. The spin-change
cross sections are calculated from the singlet and triplet phase shifts by%

\begin{equation}
\sigma_{sc}^{{}}=\frac{\pi}{k^{2}}\sum\limits_{l=0}^{\infty}{(2l+1)\sin
^{2}(\delta_{l}^{T}-\delta_{l}^{S})}. \label{eq20}%
\end{equation}

In the limit of low energies, Eq. \ref{eq20} is transformed to $\sigma
_{sc.}^{{}}=\pi\,(a_{T}^{2}-a_{T}^{2})$ \cite{10}. In addition, this parameter
is known as spin-flip collisions lead to trap losses \cite{36}. The
zero-energy cross sections for the selected mixing atoms are 3.29010$^{+11}%
a_{0}^{2}$(=9.21310$^{-10}$m$^{2})$ and 3.06510$^{10}a_{0}^{2}$
(=8.58410$^{-11}$m$^{2})$ for singlet and triplet states of the $^{40}%
$K-$^{85}$Rb, and 4.700 10$^{+11}a_{0}^{2}$(=1.316 10$^{-09}$m$^{2})$ and
8.152 10$^{+10}$(=2.28310$^{-10}$m$^{2})$ for singlet and triplet states of
the $^{39}$K-$^{87}$Rb, respectively. Spin-charge cross sections are also
7.45910$^{+10}a_{0}^{2}$(=2.08710$^{-10}$m$^{2})$ in the $^{40}$K-$^{85}$Rb
and 2.27610$^{+10}a_{0}^{2}$ (=6.37310$^{-11}$m$^{2})$ in the $^{39}$K-$^{87}$Rb.

We assume the velocity distribution is Maxwellian characterized by a kinetic
temperature $T$ , and we define mean elastic and spin-change cross sections by%

\begin{equation}
\bar{\sigma}(T)=\frac{1}{(k_{B}T)^{2}}\int\limits_{0}^{\infty}{\sigma
(E)Ee^{(-E/k_{B}T)}dE}. \label{eq21}%
\end{equation}

Assuming equal temperatures of the each pair of isotopes in the Rb-K vapor,
the corresponding rate coefficients are given by%

\begin{equation}
\label{eq22}\Re^{S,T,SC} = \left[  {\frac{8k_{B} T}{\pi\mu}} \right]
\bar{\sigma}(T).
\end{equation}

\begin{figure}
[ptb]
\begin{center}
\includegraphics[
height=2.5356in,
width=2.5936in
]%
{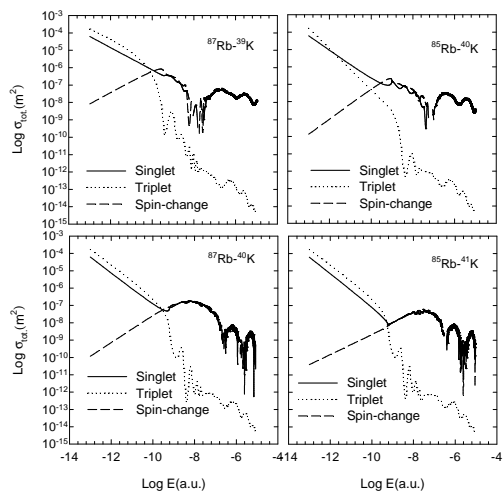}%
\caption{Cross-sections as a function of the energy in the spin charge, the
singlet and triplet states of the mixing of atoms.}%
\label{fig4}%
\end{center}
\end{figure}

We show values of the cross sections as depending on the energy in the
(m$^{2})$ in Fig. \ref{fig4} for $E$ up to 10$^{-2}$(a.u.). and list values of
the spin-change $2s_{AB}+1$ in cm$^{3}$s$^{-1}$ in the Table \ref{tab6} for
$T$ up to 1K.

Figure 4 gives cross sections for some interaction of atoms. They show that in
the low energies, less partial-wave contributes to scattering . As shown in
figure fermion-boson $^{85,87}$Rb+$^{40}$K collisions display different
properties. it may be said that $^{85}$Rb+$^{40}$K is more stability than
$^{87}$Rb+$^{40}$K, because later has  less partial waves  with decreasing
values of the energy for  both singlet and triplet state and its total cross
sections are in very low values in the same higher energy region.

\section{\textbf{CONCLUSIONS}}

The elastic scattering properties for the collision of the two mixing atoms in
a limited range of moment (k%
$<$%
$\bar{a}^{-1}\approx$0.01a.u. for Rubidium-Potassium isotopes) at low
temperatures are sensitive to the details of interaction potentials.
Scattering properties such as the scattering length, the effective range,the
phase shift and the cross section have been computed using semiclassical and a
numeric methods for the V(r) potentials as dependence of cutoff radius
adjusted by comparing with the experimental potential for ultra-cold
rubidium-potassium isotopes collision. We investigated the convergence of
these scattering properties as the depending of core radius and $K^{0}$
parameter using Quantum Defect Theory. Cross section was obtained as $\sim
$1.0$10^{-10}m^{2}$ at low energy. The phase shifts, an intermission parameter
for scattering length and effective range, has linear manner at small momenta
$k<\bar{a}^{-1}$. So, it has been determined by computing numerically and
analytically for V(r) potential. In bigger momenta than the inverse mean
scattering length, the phase shifts display parallel behavior as the
dependence of semiclassical phase shift, $\delta_{0}(k)\sim\varphi-ka$.

\begin{table}[ptbh]
\caption{Potential coefficients for Rb-K atoms in singlet and triplet state.}%
\label{tab1}%
\begin{tabular}
[c]{rrr}\hline\hline
& $^{1}\sum_{g}^{+}$ & $^{3}\sum_{u}^{+}$\\\hline
\multicolumn{1}{l}{10$^{-3}f_{1}$} &
\ \ \ \ \ \ \ \ \ \ \ \ \ \ \ \ \ \ \ \ \ 345\ \ \  &
\ \ \ \ \ \ \ \ \ \ \ \ \ \ -345\\
\multicolumn{1}{l}{10$^{-3}f_{2}$} & \ \ \ \ 1.0 \ \  & -1.0\\
\multicolumn{1}{l}{10$^{-3}f_{3}$} & \ \ \ \ 1.0 \ \  & -1.0\\
\multicolumn{1}{l}{10$^{-3}$J$_{1}^{1}$} & \ \ \ \ 5.0 \ \  & 5.0\\
\multicolumn{1}{l}{10$^{-4}$J$_{2}^{1}$} & \ \ \ 4.0 \ \  & 4.0\\
\multicolumn{1}{l}{10$^{-3}$J$_{3}^{1}$} & \ \ 7.0 \ \  & 7.0\\
\multicolumn{1}{l}{$\alpha^{1}$} & 5.256 \ \  & 5.256\\
\multicolumn{1}{l}{$\beta^{1}$} & 1.119 \ \  & 1.119\\
\multicolumn{1}{l}{10$^{-4}$C$_{6}^{2,3}$} & 1.502 \ \  & 4.274\\
\multicolumn{1}{l}{10$^{-5}$C$_{8}^{2,4}$} & 5.922 \ \  & 5.922\\
\multicolumn{1}{l}{10$^{-7}$C$_{10}^{2}$} & 6.726 \ \  & 6.726\\\hline\hline
\multicolumn{3}{l}{$^{1}$ Ref.\cite{24}, $^{2}$Ref.\cite{26}.}\\
\multicolumn{3}{l}{$^{3}$The fit constant defined in the text is$C_{6}%
^{S}=3.514$}\\
\multicolumn{3}{l}{$^{4}$The fit constant defined in the text is $C_{8}%
^{S,T}=0.865$}%
\end{tabular}
$\ $\end{table}

\clearpage

\begin{table}[ptbh]
\caption{Scattering lengths $a$ (in atomic units) for ultracold
rubidium-potassium mixtures in singlet and triplet states.}%
\label{tab2}%
\qquad%
\begin{tabular}
[c]{ccccccccccc}\hline\hline
&  & \multicolumn{3}{c}{$^{39}$K} & \multicolumn{3}{c}{$^{40}$K} &
\multicolumn{3}{c}{$^{41}$K}\\\hline
&  & $a^{i}$ & $a^{ii}$ & $a^{iii}$ & $a^{i}$ & $a^{ii}$ & $a^{iii}$ & $a^{i}$
& $a^{ii}$ & $a^{iii}$\\\hline
\multicolumn{1}{l}{$^{85}$Rb} & \multicolumn{1}{l}{$S$} &
\multicolumn{1}{l}{136.73 \ } & \multicolumn{1}{l}{136.76 \ } &
\multicolumn{1}{l}{$\_$ \ \ \ } & \multicolumn{1}{l}{161.84 \ \ } &
\multicolumn{1}{l}{161.807 \ \ } & \multicolumn{1}{l}{$\_\ \ $%
\ \ \ \ \ \ \ \ \ \ } & \multicolumn{1}{l}{212.56 \ \ } &
\multicolumn{1}{l}{212.55 \ } & \multicolumn{1}{l}{$\_$%
\ \ \ \ \ \ \ \ \ \ \ \ }\\
\multicolumn{1}{l}{} & \multicolumn{1}{l}{$T$} & \multicolumn{1}{l}{56.983} &
\multicolumn{1}{l}{56.952} & \multicolumn{1}{l}{58$_{-6}^{14}$} &
\multicolumn{1}{l}{-49.364} & \multicolumn{1}{l}{-49.390} &
\multicolumn{1}{l}{-38$_{-17}^{57}$} & \multicolumn{1}{l}{271.568} &
\multicolumn{1}{l}{306.016} & \multicolumn{1}{l}{329$_{-55}^{1000}$}\\
\multicolumn{1}{l}{$^{87}$Rb} & \multicolumn{1}{l}{$S$} &
\multicolumn{1}{l}{-3.121} & \multicolumn{1}{l}{-3.197} &
\multicolumn{1}{l}{$\_$} & \multicolumn{1}{l}{23.927} &
\multicolumn{1}{l}{23.926} & \multicolumn{1}{l}{$\_$} &
\multicolumn{1}{l}{54.459} & \multicolumn{1}{l}{54.456} &
\multicolumn{1}{l}{$\_$}\\
\multicolumn{1}{l}{} & \multicolumn{1}{l}{$T$} & \multicolumn{1}{l}{36.854} &
\multicolumn{1}{l}{36.894} & \multicolumn{1}{l}{31$_{-6}^{16}$} &
\multicolumn{1}{l}{-200.05} & \multicolumn{1}{l}{-200.04} &
\multicolumn{1}{l}{-261$_{-159}^{170}$} & \multicolumn{1}{l}{167.29} &
\multicolumn{1}{l}{167.28} & \multicolumn{1}{l}{163$_{-12}^{57}$%
}\\\hline\hline
\multicolumn{11}{l}{$^{i}$Calculated using semiclassical method for the $V(r)$
potential.}\\
\multicolumn{11}{l}{$^{ii}$Calculated using Numerov method for the $V(r)$
potential.}\\
\multicolumn{11}{l}{$^{iii}$Ref.[14], scattering lengths have been measured
with help of \ \ the conventional}\\
\multicolumn{11}{l}{cross-dimensional thermalization technique.}%
\end{tabular}
\end{table}\ \ \ \ 

\begin{description}
\item \clearpage
\ \ \ \ \ \ \ \ \ \ \ \ \ \ \ \ \ \ \ \ \ \ \ \ \ \ \ \ \ \ \begin{table}[ptbh]
\caption{Mean scattering lengths $a$ (in atomic units) and semiclassical phase
shifts for ultracold rubidium-potassium mixtures in singlet and triplet
states.}%
\label{tab3}%
\begin{tabular}
[c]{cccccccc}\hline\hline
\multicolumn{2}{c}{} & \multicolumn{2}{c}{$^{39}$K} &
\multicolumn{2}{c}{$^{40}$K} & \multicolumn{2}{c}{$^{41}$K}\\\hline
$^{85}$Rb & $a_{s},\varphi$ & 93.65 & 339.254 & 94.056 & 342.202 & 94.449 &
345.072\\
& $a_{t},\varphi$ & 68.393 & 101.089 & 68.690 & 101.967 & 68.977 & 102.823\\
$^{87}$Rb & $a_{s},\varphi$ & 93.82 & 340.486 & 94.230 & 343.467 & 94.627 &
346.369\\
& $a_{t},\varphi$ & 68.517 & 101.357 & 68.817 & 102.244 & 69.107 &
103.108\\\hline\hline
\end{tabular}
\end{table}
\end{description}

\bigskip

\begin{table}[ptbh]
\caption{Effective ranges $r_{e}$ (in atomic units) for rubidium-potassium
atom scattering.}%
\label{tab4}%
\begin{tabular}
[c]{cccccccc}\hline\hline
\multicolumn{2}{c}{} & \multicolumn{2}{c}{$^{39}$K} &
\multicolumn{2}{c}{$^{40}$K} & \multicolumn{2}{c}{$^{41}$K}\\\hline
\multicolumn{2}{c}{} & $r_{e}^{i}$ & $r_{e}^{ii}$ & $r_{e}^{i}$ & $r_{e}^{ii}$
& $r_{e}^{i}$ & $r_{e}^{ii}$\\\hline
$^{85}$Rb & $S$ & 155.323 & 155.321 & 140.841 & 140.841 & 139.505 & 139.502\\
& $T$ & 295.498 & 295.400 & 1534.41 & 1534.38 & 124.996 & 124.989\\
$^{87}$Rb & $S$ & 511.371 & 511.387 & 6637.97 & 6635.634 & 983.868 & 983.527\\
& $T$ & 838.647 & 838.413 & 386.47 & 386.535 & 103.87 & 102.995\\\hline\hline
\multicolumn{8}{l}{$^{i}$Calculated using semiclassical method for $V(r)$.}\\
\multicolumn{8}{l}{$^{ii}$Calculated using Numerov method for $V(r)$.}%
\end{tabular}
\end{table}

\begin{table}[ptbh]
\caption{Cross sections (in the $m^{2}$ unit ) and phase shifts calculated for
V(r) potential using numerical Numerov method.}%
\label{tab5}%
\begin{tabular}
[c]{llllllllll}\hline\hline
&  &  &  &  &  & \multicolumn{2}{l}{$^{x}$Rb-$^{y}$K} &  & \\\hline
& Lg$_{10}$(E) & L & Factor & $^{85-39}$ & $^{85-40}$ & $^{85-41}$ &
$^{87-39}$ & $^{87-40}$ & $^{87-41}$\\\hline
$\delta_{0}$ & -14.99 & 2 & 10$^{-3}$ & 5.17 & 4.93 & 5.86 & 3.28 & 4.48 &
2.69\\
&  & 4 & 10$^{-3}$ & 5.50 & 5.57 & 6.61 & 3.64 & 5.36 & 3.36\\
& -11.57 & 2 & $-$ & 0.256 & 0.245 & 0.289 & 0.165 & 0.223 & 0.136\\
&  & 4 & $-$ & 0.271 & 3.261 & 0.322 & 1.562 & 0.265 & 0.169\\
$d\sigma$ & -14.99 & 2 & 10$^{-12}$ & 5.00 & 4.92 & 7.67 & 2.24 & 4.73 &
1.58\\
&  & 4 & 10$^{-12}$ & 3.63 & 3.55 & 5.61 & 1.62 & 3.43 & 1.14\\
& -11.08 & 2 & 10$^{-12}$ & 4.57 & 4.52 & 6.63 & 2.15 & 4.33 & 1.53\\
&  & 4 & 10$^{-12}$ & 3.32 & 3.26 & 4.85 & 1.56 & 3.14 & 1.12\\\hline\hline
\end{tabular}
\end{table}

\clearpage

\begin{table}[ptbh]
\caption{Singlet, Triplet and Spin-change rate coefficients for different
isotops of the mixed Rb-K. The numbers in brackets denote multiplicative
powers. a) $^{85}$Rb-$^{39}$K, b)$^{85}$Rb-$^{40}$K, c)$^{85}$Rb-$^{41}$K,
d)$^{87}$Rb-$^{39}$K, e)$^{87}$Rb-$^{40}$K, f) $^{87}$Rb-$^{41}$K.}%
\label{tab6}
\begin{tabular}
[c]{cccccccc}\hline\hline
log$_{10}$(K)(a) & R$_{S}$(m$^{3}$s$^{-1})$ & R$_{T}$(m$^{3}$s$^{-1})$ &
R$_{SC}$(m$^{3}$s$^{-1})$ & log$_{10}$(K)(d) & R$_{S}$(m$^{3}$s$^{-1})$ &
R$_{T}$(m$^{3}$s$^{-1})$ & R$_{SC}$(m$^{3}$s$^{-1})$\\\hline
-8 & 2.00(-12) & 5.48(-12) & 1.30(-16) & -8 & 2.02(-12) & 5.50(-12) &
2.72(-16)\\
-7 & 1.13(-12) & 3.09(-12) & 7.34(-17) & -7 & 1.12(-12) & 3.05(-12) &
1.51(-16)\\
-6 & 6.77(-14) & 1.80(-13) & 1.41(-15) & -6 & 6.82(-14) & 1.78(-13) &
1.89(-15)\\
-5 & 4.75(-14) & 8.57(-14) & 1.45(-14) & -5 & 5.23(-14) & 8.47(-14) &
1.95(-14)\\
-4 & 5.33(-14) & 9.27(-15) & 5.17(-14) & -4 & 4.75(-14) & 9.13(-15) &
5.56(-14)\\
-3 & 1.56(-13) & 4.19(-16) & 1.74(-13) & -3 & 2.08(-14) & 4.13(-16) &
2.64(-14)\\
-2 & 3.71(-13) & 1.59(-17) & 4.31(-13) & -2 & 3.05(-14) & 1.56(-17) &
3.53(-14)\\
-1 & 5.45(-13) & 1.44(-18) & 6.36(-13) & -1 & 1.29(-13) & 1.42(-18) &
1.50(-13)\\
0 & 6.52(-13) & 4.35(-19) & 7.61(-13) & 0 & 1.84(-13) & 4.32(-19) &
2.15(-13)\\\hline
log$_{10}$(K)(b) & R$_{S}$(m$^{3}$s$^{-1})$ & R$_{T}$(m$^{3}$s$^{-1})$ &
R$_{SC}$(m$^{3}$s$^{-1})$ & log$_{10}$(K)(e) & R$_{S}$(m$^{3}$s$^{-1})$ &
R$_{T}$(m$^{3}$s$^{-1})$ & R$_{SC}$(m$^{3}$s$^{-1})$\\\hline
-8 & 2.01(-12) & 5.53(-12) & 4.39(-18) & -8 & 2.00(-12) & 5.48(-12) &
1.30(-16)\\
-7 & 1.09(-12) & 2.99(-12) & 2.37(-18) & -7 & 1.13(-12) & 3.09(-12) &
7.34(-17)\\
-6 & 6.40(-14) & 1.75(-13) & 1.83(-16) & -6 & 6.77(-14) & 1.80(-13) &
1.41(-15)\\
-5 & 3.28(-14) & 8.33(-14) & 2.22(-15) & -5 & 4.75(-14) & 8.57(-14) &
1.45(-14)\\
-4 & 1.97(-14) & 8.92(-15) & 1.63(-14) & -4 & 5.33(-14) & 9.27(-15) &
5.17(-14)\\
-3 & 2.47(-14) & 4.03(-16) & 2.83(-14) & -3 & 1.56(-13) & 4.19(-16) &
1.74(-13)\\
-2 & 1.16(-14) & 1.52(-17) & 1.34(-14) & -2 & 3.71(-13) & 1.59(-17) &
4.31(-13)\\
-1 & 7.75(-14) & 1.39(-18) & 9.06(-14) & -1 & 5.45(-13) & 1.44(-18) &
6.36(-13)\\
0 & 9.56(-14) & 4.24(-19) & 1.12(-13) & 0 & 6.52(-13) & 4.35(-19) &
7.61(-13)\\\hline
log$_{10}$(K)(c) & R$_{S}$(m$^{3}$s$^{-1})$ & R$_{T}$(m$^{3}$s$^{-1})$ &
R$_{SC}$(m$^{3}$s$^{-1})$ & log$_{10}$(K)(f) & R$_{S}$(m$^{3}$s$^{-1})$ &
R$_{T}$(m$^{3}$s$^{-1})$ & R$_{SC}$(m$^{3}$s$^{-1})$\\\hline
-8 & 2.03(-12) & 5.58(-12) & 1.31(-18) & -8 & 1.96(-12) & 5.36(-12) &
1.27(-16)\\
-7 & 1.05(-12) & 2.88(-12) & 6.76(-19) & -7 & 1.11(-12) & 3.03(-12) &
7.19(-17)\\
-6 & 6.23(-14) & 1.72(-13) & 4.89(-18) & -6 & 6.63(-14) & 1.76(-13) &
1.38(-15)\\
-5 & 2.94(-14) & 8.10(-14) & 6.08(-17) & -5 & 4.65(-14) & 8.39(-14) &
1.42(-14)\\
-4 & 3.98(-15) & 8.59(-15) & 8.95(-16) & -4 & 5.22(-14) & 9.08(-15) &
5.06(-14)\\
-3 & 1.14(-14) & 3.88(-16) & 1.20(-14) & -3 & 1.52(-13) & 4.11(-16) &
1.71(-13)\\
-2 & 4.53(-14) & 1.45(-17) & 5.21(-14) & -2 & 3.64(-13) & 1.55(-17) &
4.22(-13)\\
-1 & 2.49(-14) & 1.36(-18) & 2.91(-14) & -1 & 5.33(-13) & 1.41(-18) &
6.23(-13)\\
0 & 1.50(-14) & 4.03(-19) & 1.72(-14) & 0 & 6.39(-13) & 4.26(-19) &
7.45(-13)\\\hline\hline
\end{tabular}
\end{table}

\clearpage
%

\bigskip\clearpage

\clearpage
\bigskip\qquad%

\clearpage
%

\clearpage

\end{document}